# Electrical excitation of color centers in phosphorus-doped diamond Schottky diodes

Florian Sledz[1,*], Igor A. Khramtsov[2], Assegid M. Flatae[1], Stefano Lagomarsino[3], Silvio Sciortino[3,4,5], Shannon S. Nicley[6,7], Rozita Rouzbahani[6], Paulius Pobedinskas[6], Tianxiao Guo[8], Xin Jiang[8], Paul Kienitz[9], Peter Haring Bolivar[9], Ken Haenen[6], Dmitry Yu. Fedyanin[1,2] and Mario Agio[1,5]

[1]Laboratory of Nano-Optics, University of Siegen, Germany
[2]Moscow Center for Advanced Studies, Moscow, Russia
[3]Istituto Nazionale di Fisica Nucleare, Sezione di Firenze, Italy
[4]Department of Physics and Astronomy, University of Florence, Italy
[5]National Institute of Optics (INO), National Research Council (CNR), Italy
[6]Institute for Materials Research (IMO) & IMOMEC, Hasselt University & IMEC vzw, Belgium
[7]Department of Electrical and Computer Engineering, Michigan State University, USA
[8]Lehrstuhl für Oberflächen- und Werkstofftechnologien, University of Siegen, Germany
[9]Group of Graphene-based Nanotechnology, University of Siegen, Germany

*Corresponding author: Florian.Sledz@uni-siegen.de

## Abstract
A robust quantum light source operating upon electrical injection at ambient conditions is desirable for practical implementation of quantum technologies, such as quantum key distribution or metrology. Color centers in diamond are promising candidates as they are photostable emitters at room and elevated temperatures. The possibility of their electrical excitation has already been demonstrated within p-i-n diodes. However, this requires the growth of complex diamond structures. In contrast to these conventional approaches, we demonstrate the emission of color centers under electrical pumping in a novel Schottky diode configuration based on hydrogen passivated n-type diamond, which holds promise for integrated single-photon emitting devices based on color centers in diamond.

## Introduction
Among the fundamental building blocks of optical quantum technologies, quantum-light sources play a crucial role. Indeed, most applications ranging from secure communication lines to quantum metrology and optical quantum computers [1–4], require operation at the single-photon level.

There are many known methods of generating single photons. For instance, the excitation of atoms/ions [5–7], isolated molecules [8–10], defects in semiconductors [11,12], and semiconductor quantum dots [13–15]. Although atoms and ions provide spectrally indistinguishable photons, their operation is limited due to the complicated trapping and cooling mechanisms needed to keep them in place. Molecules show blinking, limited photostability, and they cannot be reliably used for applications under ambient conditions. In the last few years, epitaxial quantum dots have shown promising results, however their operation is limited to cryogenic conditions and requires optical microcavities for efficient emission [16,17].

It has been demonstrated that color centers in diamond can produce single photons at a rate of more than $10^6$ counts per second (cps) under optical excitation [11] and they can be operated at ambient an elevated temperatures [18]. However, from a practical point of view, it is highly desirable to control them under electrical excitation to develop a scalable quantum photonic architecture [19,20]. Electrical excitation does not only allow to avoid the use of bulky lasers, but it also gives the possibility to independently manipulate hundreds and thousands of single-photon sources regardless of the separation distance among them. Thus, it becomes possible to develop a large-scale quantum photonic circuit on a single chip. However, electrical excitation of color centers in diamond is very challenging. First, diamond is a unique material between semiconductors and insulators that features a bandgap energy ($E_g$) of 5.47 eV. Therefore, it is extremely difficult to create a high density of free carriers, especially electrons, due to the high activation energy of dopants [21,22]. Second, the common approach for electrical excitation of color centers in diamond is forward biased p-i-n diodes that contain a color center in the i-type region [23–27]. Electrons injected from the n-type region and holes injected from the p-type region recombine at the color center, which results in photon emission [28,29]. However, the fabrication of a diamond p-i-n diode is much more complicated than the fabrication of such structure based on conventional semiconductors like Si and GaAs. This is mainly due to the absence of an efficient technique of creating regions with high free carrier densities by ion implantation, which is used in microelectronics for low-cost production of integrated electronic circuits [30].

Here, we propose and experimentally demonstrate a fundamentally different approach for electrical excitation of color centers in diamond, which gives the possibility to fabricate any number of independently operating single-photon emitting diodes on a single chip and therefore to develop scalable quantum photonic circuits. In contrast to the common methods based on p-n and p-i-n structures, our approach is based on a Schottky barrier diode on n-type diamond (phosphorus-doped) [31]. This dramatically simplifies the fabrication and avoids complicated processes of structuring and regrowing of diamond. Diodes of any size can be created anywhere on a chip by performing proper metallization, which reduces the fabrication cost and leads to unprecedented scalability. Although single-photon emission is not addressed in this work, our findings introduce an approach that can be readily applied for obtaining an electrically-pumped single-photon source.

## Device concept and numerical simulations

Schottky diodes are known as majority carrier devices, i.e., the density of one type of free charge carriers (electrons and holes) is many orders of magnitude lower than the density of the other type at any point of the device at any bias voltage. Thus, the electron-hole recombination rate at the color center should be zero in Schottky diodes, which does not allow to efficiently excite the color centers electrically. Although it is still possible to observe electroluminescence in such devices due to impact excitation, it is less efficient and usually requires large bias voltages of hundreds of volts [32,33]. Nevertheless, in the 1960s, it was theoretically shown that minority carriers can be injected directly from the metal to the semiconductor, if the Schottky barrier height (SBH) is comparable with or higher than the bandgap of the semiconductor [34]. The problem is that almost no material satisfies this requirement [35] due to the surface states at the metal-semiconductor interface, which pin the Fermi level [36] and have a large influence on the achievable barrier height.

Recently, it was theoretically predicted that hydrogen termination of the diamond surface can dramatically reduce the density of surface states [37] and therefore their influence on the barrier height vanishes. Since the electron affinity of diamond is negative (down to $\chi_e = -2.01$ eV) [38], the SBH can almost be equal to or higher than the bandgap energy of diamond, if the work function of the metal is sufficiently high. In this work, we use gold, since it features one of the largest work functions (up to $\Phi_M = 5.3$ eV) [39]. The theory predicts the SBH between gold and hydrogen terminated n-type diamond, $\Phi_{B,e}$, to be higher than 4.95 eV [37]. Accordingly, the SBH for holes $\Phi_{B,h} = E_g - \Phi_{B,e}$ is less than 0.5 eV. Such a low potential barrier for holes gives the possibility to inject a significant density of holes, which are minority carriers in n-type diamond, directly from gold under high forward bias. Hence, high densities of both electrons and holes can be created near the Schottky contact, which can recombine at the color center at a high rate and efficiently excite it. A band structure of such a Schottky junction is shown in Fig. 1a in thermal equilibrium (top) and under forward bias (bottom). As electrons and holes are present in the vicinity of a color center, they can pump it into an excited state from where it can return back to the ground state by the emission of a photon [28].

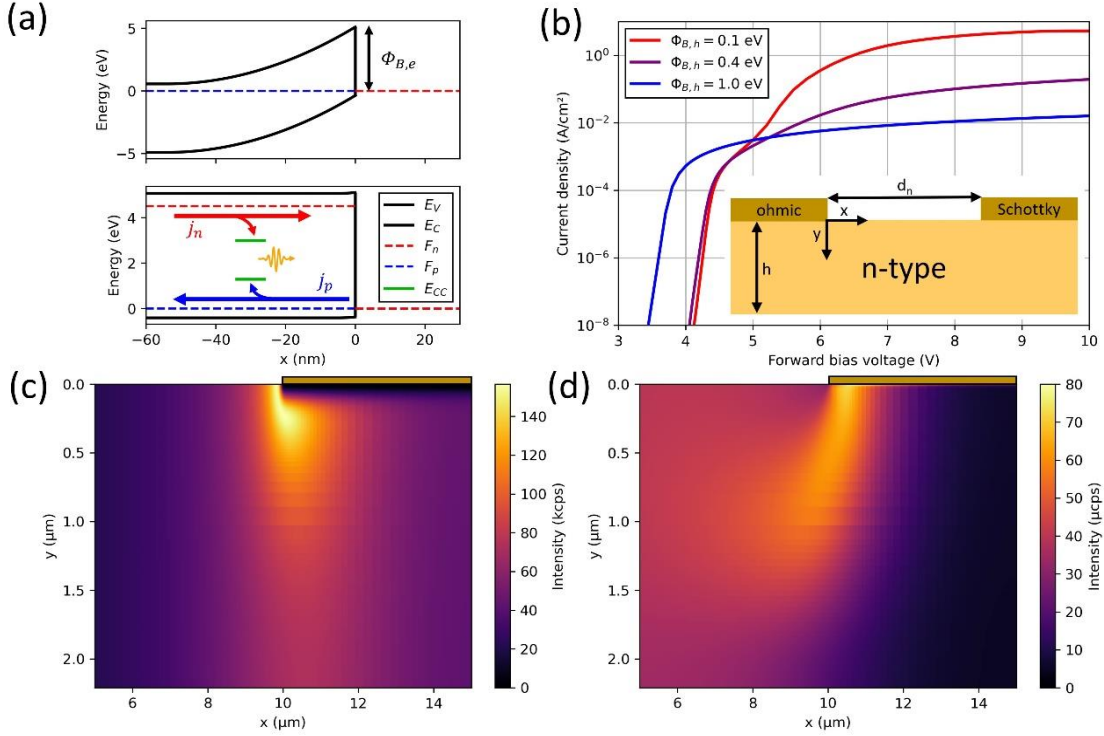

Figure 1. (a) Schematic energy diagram of a Schottky junction with a barrier height of 5.1 eV in equilibrium (top) and forward bias voltage of 4.5 V (bottom). The black lines show the valence ($E_V$) and conduction bands ($E_C$), the red and blue dashed lines show the positions of the quasi-Fermi levels for electrons ($F_n$) and holes ($F_p$), respectively. The blue and red arrows (bottom) show the direction of the electron ($j_n$) and hole current densities ($j_p$). A small portion of the charge carriers are captured by color centers (energy levels $E_{CC}$ shown in green), which leads to the emission of a photon (orange). (b) Simulated current voltage characteristic for different Schottky barrier heights for holes, $\Phi_{B,h}$. For $\Phi_{B,h}$ below 0.4 eV, an effective injection of holes can be seen as a step in the IV curve at ~5 V. The inset shows the schematic of the simulated structure. The coordinate system is fixed to the edge of the ohmic contact. Details can be found in the text. (c, d) Single-photon emission rates for an ideal source with 100% quantum efficiency in proximity to a Schottky contact with a bias voltage of 10 V and barrier heights for holes of 0.1 and 1.0 eV, respectively. The position of the Schottky contacts is marked with the boxes above of the color maps.

Self-consistent numerical simulations of the proposed structure have been performed with the software Silvaco TCAD. For the simulation, a free-standing n-type (phosphorus-doped) diamond film with a thickness ($h$) of 2.2 μm, a high donor concentration of $10^{18}$ cm$^{-3}$, a compensation ratio of 10 % and mobilities of $\mu_n = \mu_p = 390 \frac{\text{cm}^2}{\text{Vs}}$ were used to consider the high doping level of the thin film [40]. On top of the diamond an ohmic and a Schottky contact are placed with a spacing ($d_n$) of 10 μm in between, which are defined as boundary conditions at the edge of the simulation region. The simulations show an effective injection of holes by thermionic emission under large bias voltages. If the barrier height for holes is smaller than 0.4 eV, the hole current can even surpass the electron current. This can be seen as a second step in the IV-curves in Fig. 1b. At room temperature, the charge carrier capture is the limiting process of the luminescence dynamics, with capture rates of $c_n n$ and $c_p p$ less than $10^6$ s$^{-1}$. Here $c_n = 1.7 \times 10^{-8}$ cm$^3$s$^{-1}$ and $c_p = 3.9 \times 10^{-7}$ cm$^3$s$^{-1}$ are the capture rate constants for electrons and holes, respectively. According to references [28,29], the single-photon emission rates can then be calculated from the charge carrier densities $n$ and $p$ for electrons and holes, respectively, by:

$$R = \varphi \left( \frac{1}{c_n n} + \frac{1}{c_p p} \right)^{-1}, \qquad (1)$$

here $\varphi$ is the quantum efficiency of the respective color center. Figs. 1c and 1d show the single-photon emission rates for a color center with unity quantum efficiency ($\varphi = 1$) for Schottky barrier heights for holes of 0.1 eV and 1.0 eV at a forward bias voltage of 10 V, which leads to total current densities of $j = 5.3 \text{ A/cm}^2$ and $j = 1.6 \times 10^{-2} \text{ A/cm}^2$, respectively. Under these conditions, maximal emission rates of 150 kcps and 60 μcps are expected from an emitter with $\varphi = 1$ near the Schottky junction for the different configurations. While the first emission rate is comparable with optical excitation rates [41], the latter value is much lower than typical dark counts and noise levels, and it makes the observation of single-photon emission at such large barrier heights practically impossible. Additionally, one must consider the quantum efficiencies of the color centers, and the collection and detection efficiencies of the experimental setup, such that less than 1% of the excitation results in a detected photon during the experiment.

# Sample preparation

### A. Diamond growth

A phosphorus-doped homoepitaxial diamond film was grown on top of the surface of a (111) oriented 2.5 mm × 2.5 mm × 0.5 mm high-pressure high-temperature (HPHT) substrate (Sumitomo) by microwave plasma enhanced chemical vapor deposition (MW PE CVD) in a 2.45 GHz ASTeX PDS17 CVD reactor. During the growth, a total flow rate of 500 sccm and a pressure of 140 Torr were kept constant. The hydrogen ($H_2$) rich plasma contained 0.15% of methane ($CH_4$). For the incorporation of phosphorus, phosphine ($PH_3$) was introduced into the plasma with a constant ratio of $PH_3/CH_4$= 5000 ppm. The substrate temperature was maintained at 1000 °C by variation of the applied microwave power between 1.0 and 1.3 kW. The $H_2$ and $CH_4$ gasses are filtered to a purity of <1 ppb, while the $PH_3$ is obtained from a source diluted to 200 ppm in $H_2$. After a growth time of eight hours, a n-type layer with a thickness of 2.2 ± 0.2 μm was obtained. The sample thickness was measured by a Mitutoyo Linear Gage (Model 542–158) [42].

### B. Color center creation

Silicon ion implantation allows to create silicon-vacancy (SiV) color centers and vacancies for nitrogen-vacancy (NV) color centers. Thermal annealing enables the activation of both centers. At first, the Si ions were implanted in the first 200 nm of the n-type layer by ion-beam implantation with different ion doses ($5 \times 10^7$, $10^{12}$, $10^{13}$ and $10^{14}$ cm$^{-2}$). The Si-ion implantation is based on a 3 MV Tandetron accelerator equipped with a HVEE860 negative sputter ion source. Based on the charge state and terminal voltage the Si ion species ($Si^+$, $Si^{2+}$ and $Si^{3+}$) can be accelerated to energies up to 10 MeV [43]. Two aluminum foils with a total thickness of 4.6 μm have been used to reduce the ion energy down to a few tens of keV to allow for the desired shallow implantation. Utilizing a home-built furnace, the sample was thermally annealed at a temperature of 1200 °C for ~1 h under high-vacuum conditions (~10$^{-7}$ mbar) to enable the activation of the color centers. Beside the activation of SiV color centers, the created vacancies also recombine with incorporated N atoms during thermal annealing. This results in the formation of NV color centers within the n-type diamond layer. Details on the optical studies of the color centers in the n-type sample can be found in a previous publication [41].

### C. Device fabrication

Before device fabrication, the sample was cleaned in acetone and isopropanol at ~50 °C to remove organic contamination from its surface. In the first fabrication step, a lithographic mask was created on the cleaned diamond surface followed by thermal evaporation of titanium (16 nm) and gold (150 nm). The contacts were structured by a lift-off process such that a single pad was formed in the different implanted regions. Within these contact pads, holes with a diameter of 100 µm and pitch of 125 µm were left blank for the formation of the Schottky contacts.

Next, the sample was transferred to another CVD reactor (ASTeX) for hydrogen plasma treatment of the diamond surface. The conditions of the plasma treatment are listed in Table 1. The pressure was adjusted in the given bounds to stabilize the temperature to 850 °C during the treatment time.

*Table 1: Process parameters of the hydrogen plasma treatment.*

| Power | Temperature | Pressure | $H_2$ flow rate | Time |
| --- | --- | --- | --- | --- |
| 850 W | 850°C | 20-26 torr | 400 sccm | 45 min |

After treatment, the sample was left in the hydrogen atmosphere for ~2.5 h to cool down to room temperature. This plasma treatment has two effects: first the ohmic contacts are thermally annealed by the formation of titanium-carbide at these high temperatures and second, the diamond surface acquires a reduced electron affinity by the hydrogen passivation [44] and thus a lowering of the achievable barrier height for holes. The recipe was tested by the formation of a surface conductive layer on an intrinsic diamond (see supplementary information).

After cooling down, the sample was removed from the CVD reactor, and it was transferred to a thermal evaporation chamber for the deposition of the Schottky contact. For this, a 150 nm-thick gold layer was deposited over the whole sample without an additional adhesion layer, as the latter would affect the properties of the Schottky junction.

For patterning of the final devices, a complementary positive lithographic mask was aligned relative to the ohmic contacts, visible beneath the gold layer. The gold in the undesired regions was wet etched using $KI_2/I_2$ solution (Micro chemicals) for 2 min. This resulted in circular Schottky contacts with a diameter of 70 µm surrounded by a common ohmic contact with circular holes with a diameter of 100 µm (see Fig. 2a).

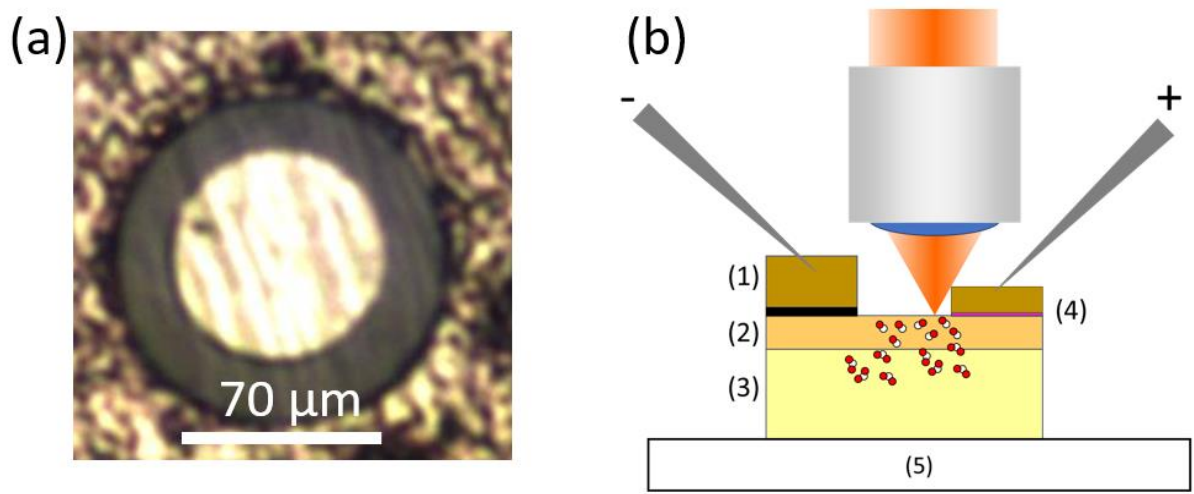

*Figure 2. (a) Microscope image of a single diamond Schottky light emitting diode. The inner disk is the Schottky contact with a diameter of 70 µm and the surrounding (rough) contact is a joint ohmic contact. (b) Illustration of the fabricated and probed diamond Schottky light emitting diode and the measurement setup. The diode consists of an ohmic contact (gold on titanium carbide) (1), the n-type diamond layer (2), the HPHT substrate (3) and the Schottky contact (gold on the hydrogen passivated*

diamond) (4). The red and white circles mark nitrogen atoms and vacancies in the different diamond layer. The sample is placed on a Peltier element (5) for heating of the sample.

## Experimental setup

The electroluminescence measurements were performed using a home-built confocal microscope setup. The setup consists of a 532 nm 5 mW laser diode (Thorlabs) for optical excitation and the alignment of the device in the microscope. The laser is coupled to the collection path by a 10/90 beam-splitter (Semrock) and focused by the microscope objective (Zeiss LD C Epiplan-APOCHROMAT 50x, 0.6 NA) onto the sample surface. The collected light is either sent to a CMOS camera (Andor Zyla 4.2Plus) for widefield imaging or it is coupled into a multimode fiber connected to a spectrometer (OceanOptics Flame 2000). Optical excitation is used for localizing the emitters for electrical excitation and detection.

The diamond sample with fabricated devices on the surface was placed on top of a home-built sample holder with a Peltier heater, allowing for the thermal activation of free electrons in the n-type layer by heating the sample up to 150 °C during the experiment. The Schottky and ohmic contacts were connected to a source measure unit (Keithley 2450) via two piezo micro probers (Imina Technologies miBots), which allow to address all devices independently. Here the cathode is placed on a joint ohmic contact pad for different devices in one implanted region, while the anode is placed on the center of the Schottky contact. Fig. 2b shows a sketch of this approach.

## Results and discussion

### A. IV-Characteristics

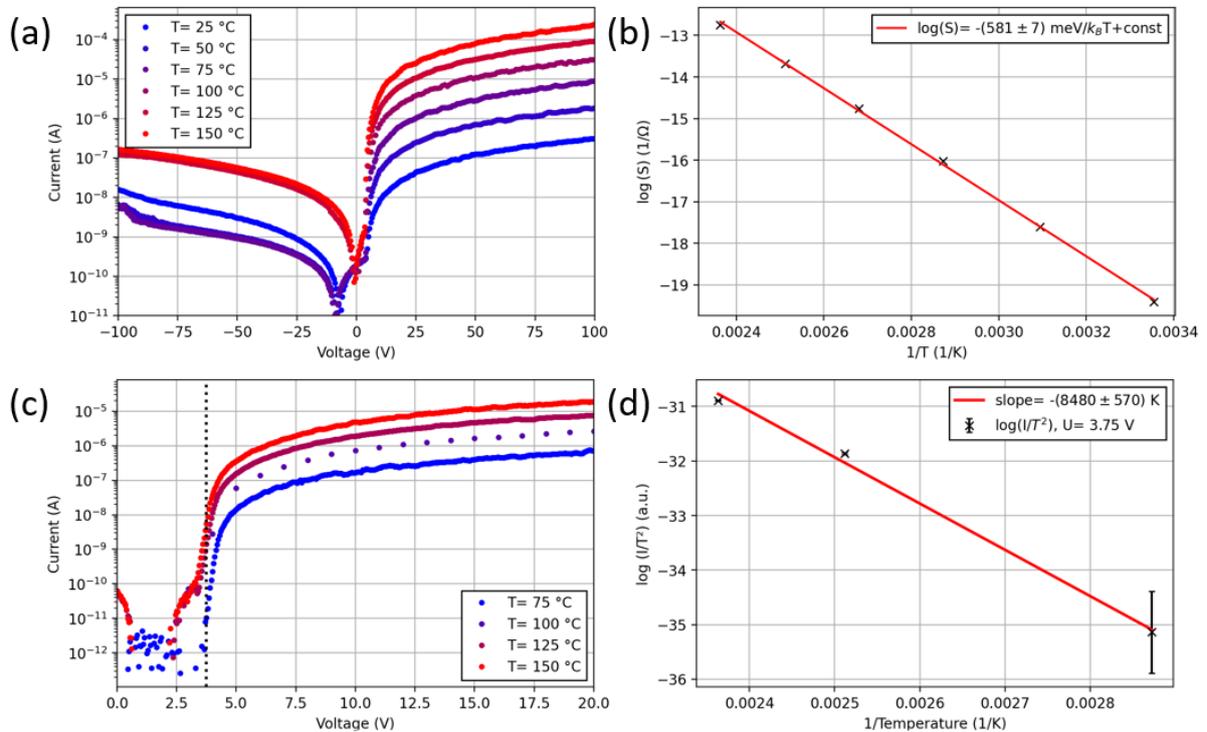

*Figure 3. (a) Raw IV-curves measured at different temperatures. The reverse bias current is dominated by the parallel current through the measurement devices. (b) Natural logarithm of the on-conductance (S) as a function of inverse temperature. (c) IV curves corrected for a parallel resistance of 22.9 GΩ. The black dotted line is located at 3.75 V, at this voltage the current is read for the calculation of the barrier height. (d) Temperature dependence of the current over temperature squared as a function of inverse temperature. From the slope, a barrier height for electrons of $\Phi_{B,e} = (4.48 \pm 0.05)$ eV is determined.*

The IV-curve are recorded in the region implanted with a Si dose of $10^{12}$ cm$^{-2}$ at temperatures in the range from 25 °C to 150 °C. The measurements show a clear diode behavior with a highly temperature dependent forward bias current and a reverse current, which is mainly given by the parallel currents through the measurement devices (see Fig. 3a). The shift of the zero crossing of the current at lower voltages is given by the discharge of an internal capacitor of the measuring device during the experiment. At a temperature of 150 °C, a rectification ratio of ~1000 (maximal forward current divided by maximal reverse current) is observed with no sign of an electrical breakdown in the measured voltage range. To characterize the temperature dependence of the forward bias current, the on-state conductance is calculated as the average slope of the IV-curve above 75 V. The conductance ($S$) is plotted in Fig. 3b in a linearized form, by calculating the natural logarithm, as a function of the inverse temperature. A linear function is fitted to this data resulting in the slope of $-\frac{(581 \pm 7)\,\text{meV}}{k_B T}$, where $k_B$ is the Boltzmann constant and $T$ the temperature, which agrees well with the activation energy of phosphorus in the n-type diamond [22]. The on-state resistance reduces from $\approx 220\,\text{M}\Omega$ at room temperature down to $\approx 360\,\text{k}\Omega$ at 150 °C. As only direct current measurements have been done, the barrier height is estimated from the temperature dependence of the thermionic emission current [45]. Here the approximation of an ideal diode behavior is made, resulting in the current dependence at $U = 3.75\,\text{V} \gg \frac{k_B T}{q}$:

$$\log\left(\frac{I}{T^2}\right) = const - \frac{q(\Phi_{B,e} - 3.75\,\text{V})}{k_B T}, \quad (2)$$

such that the barrier height can be calculated from the slope obtained by a linear fit to the data obtained from Fig. 3c (black dotted line at 3.75 V). A barrier height for electrons of $\Phi_{B,e} = (4.48 \pm 0.05)\,\text{eV}$ was determined, which corresponds to a barrier height for holes of $\Phi_{B,h} \approx E_g - \Phi_{B,e} = (0.97 \pm 0.05)\,\text{eV}$. The value of the barrier height for electrons is still close to previous results of Suzuki et al., where constant values of 4.3 eV have been observed for several metals due to Fermi-level pinning by surface states [46]. This indicates that the removal of surface states needs further improvement.

### B. Electroluminescence studies

The large barrier height for holes does not allow the observation of bright luminescence of color centers. The electroluminescence of ensembles of NV color centers is visible even by bare eye at elevated temperatures and other electroluminescence signals due to H3 color centers (also named NVN) are detected even at room temperature (see supplementary information). The signal of the SiV color centers is not distinguishable from the phonon sidebands of the stronger NV emission in this sample. In higher silicon-implanted regions, the typical zero-phonon line (ZPL) of SiV color centers is visible on top of the tail of other emission bands (see supplementary information).

Wide-field imaging of the electroluminescence signals as a function of the applied voltage allows for a direct spatial analysis of the emission characteristics across the whole device. Fig. 4a shows a wide-field image at an applied voltage of 100 V at 150 °C. The image shows a radially decaying intensity concentrated at the inner Schottky contact reaching peak values of ~63000 counts with an integration time of 0.5 s. The intensity at the positions of the three white lines (x1, x2, x3) in Fig. 4a is averaged over 50 pixels for a smoothing of the data to determine the onset voltage of the electroluminescence at three different locations with increasing distance from the Schottky contact. The determined intensities are plotted in Fig. 4b as a function of the applied voltage (colored lines), along with the current through the whole device (black line). All the intensities show an onset at ~4 V indicated by the black dotted line, which is defined by crossing a stable signal above 1 count as a threshold. At this point one can also see that the IV-curve starts to behave more ohmic, which indicates a high voltage drop across the Schottky junction and large band bending. At large applied bias voltages, drift currents determine the local charge carrier densities proportional to the voltage. This leads to an almost linear dependence between the applied voltage and the measured electroluminescence signals. As the diode exhibits an ohmic behavior in this regime, the current and the signals show a linear dependence as well (see supplementary information). The emission rates are also highly dependent on the location of the color center as the local densities of the minority carriers decrease with distance from the Schottky contact, resulting in different dependencies at the different locations (x1, x2, x3) (see Fig. S3 in the supplementary information).

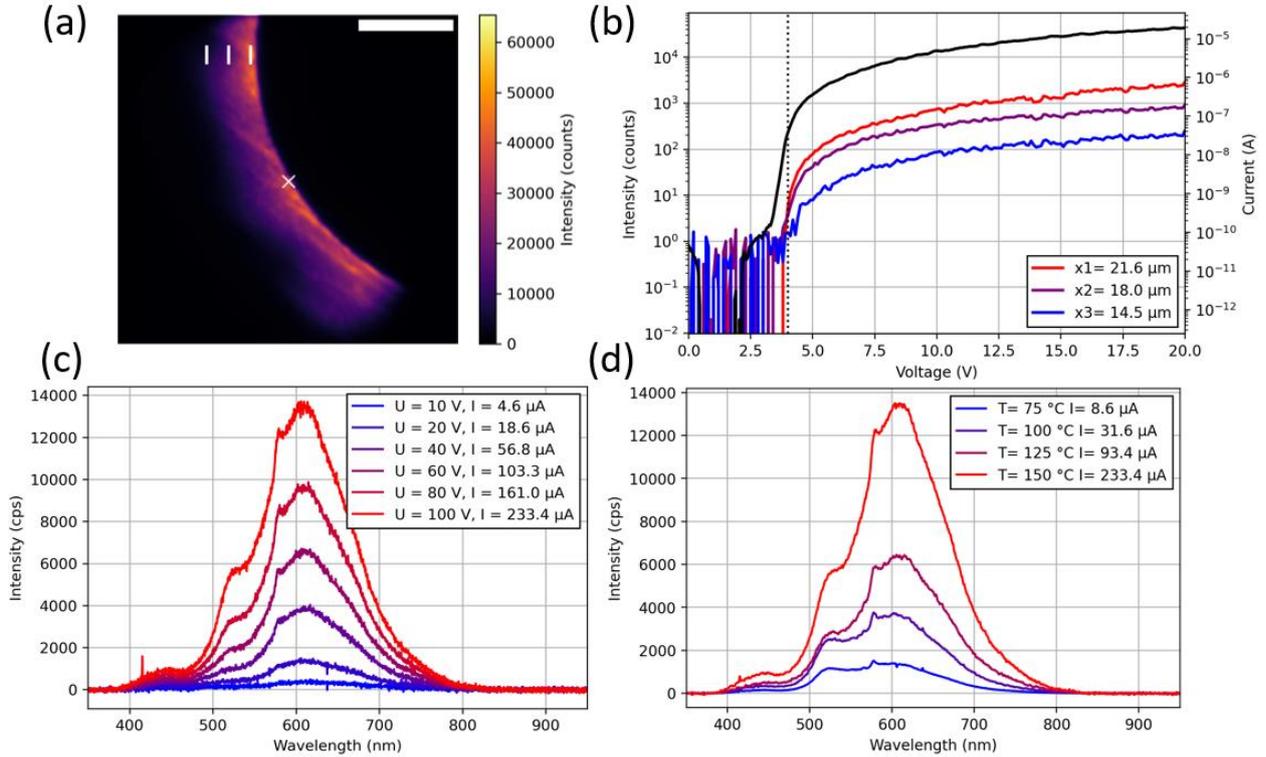

*Figure 4. (a) Widefield electroluminescence images at 150°C and applied bias voltages of 100 V. The scalebar in the top right corner corresponds to a size of 15 μm. The white × in the center marks the location of the optical axis, where the electroluminescence spectra are recorded. The three smaller vertical lines indicate the positions where the intensities in (b) are obtained as an average over 50 pixels. (b) Intensities (colored) across the white lines marked in (a) and the current through the whole device (black). Electroluminescence is visible above a turn-on voltage of 4 V (indicated by the black-dotted vertical line). At this point, the IV-curve starts to become ohmic, which indicates a high potential difference across the Schottky barrier. (c) Electroluminescence spectra recoded at different applied bias voltages for a temperature of 150 °C. The center of the collection area is marked by the white × in (a). The spectra show a dominant emission of $NV^0$ color centers along with the emission of H3 color centers. (d) Smoothed electroluminescence spectra at an applied bias voltage of 100 V at different temperatures. Due to the increase of temperature, the number of electrons in the n-type diamond increases and leads to a stronger emission of the NV color centers compared to the emission of the H3 color centers.*

The electroluminescence spectra shown in Figs. 4c and 4d are recorded at the white × marked in Fig. 4a, which indicates the optical axis of the confocal microscopy setup. The spectra are acquired in an automatic loop by changing the applied forward bias voltage, recording current and spectra at every step with an integration time of 2 s. The spectra at 150 °C and different applied bias voltages (see Fig. 4c) show a dominant emission of neutral-charge state NV ($NV^0$) color centers, with a ZPL at ~580 nm and a broad phonon-sideband centered at 620 nm. The emission of the negative-charge state of the NV color centers is not observed. The reason is that a single captured electron cannot provide enough energy to transform the color center to the negative charge state and bring it at the same time to the excited state [28].

Along with the emission of the NV color centers, two other emission bands are observed. The first one is centered at 430 nm and can be associated to band-A emission [47]. The second one is centered at

~510 nm and is related to H3 color centers [47]. The observed electroluminescence is the sum of the three color centers' emission spectra, which are visible as different plateaus. While the emission of the $NV^0$ color centers increases proportional to the current through the device, the emission of the H3 color centers is getting stronger with an increase in the applied voltage. This can be seen by the changes of the spectral shape at different applied bias voltages in Fig. 4c and more clearly from the different composition of the spectra at different temperatures in Fig. 4d. This could be either due to a different excitation scheme of these color centers or a significant amount of the charge carriers is moving through the nitrogen rich HPHT substrate and not only the n-type thin film. The latter can be caused by the lateral geometry of the structure leading to a downstream of holes into the substrate from the edge of the Schottky contact. A reduction of the spacing between the contacts down to 1 µm or less increases the lateral electric field within the n-type layer, which can confine the charge carriers into the doped film. Fig. 4d shows that the electroluminescence signal of the NV color centers increases stronger with the temperature compared to the other emission bands. This agrees with the predictions of Fedyanin and Agio [28] that the increase of charge carriers (mainly electrons) compensates the reduction in the quantum efficiency. Thus, the excitation of the NV color centers is limited by the electron capture rather than the hole capture, which should be seen by a saturation behavior of the emission as a function of current. Besides the needed increase in the barrier height for a stronger injection of holes, a further increase in the electron density is required to observe the electroluminescence of the NV color centers at room temperature.

## Conclusions

In conclusion, we have demonstrated a different approach for the electrical excitation of color centers in a n-type diamond Schottky configuration and studied its electrical properties and the electroluminescence under different conditions. Barrier heights on the order of 4.5 eV have been obtained allowing only for a small injection of holes into the diamond lattice. The observed emission is dominated by nitrogen related defects (H3 and $NV^0$). We have shown that the emission of the NV color centers increases with temperature as the charge carrier density in the n-type layer increases, which shows that electron capture is still the limiting factor in the emission process. The N-rich substrate prevented us from observing single-photon emission, but the introduction of a buffer diamond layer or substrate removal would allow one to address single color centers. Hence, this work paves the way for efficient electrically driven single-photon sources as the approach can be combined with thin diamond membranes as a host material for the single emitters [48] in combination with planar optical antennas for an efficient light extraction [49,50].

## Acknowledgements

The authors gratefully acknowledge financial support from the University of Siegen, the German Research Association (DFG) (INST 221/118-1 FUGG, 410405168), the Hasselt University Special Research Fund (BOF), the Research Foundation Flanders (FWO-G0D4920N), and the Methusalem NANO network. The authors also acknowledge INFN-CHNet, the network of laboratories of the INFN for cultural heritage, for support and precious contributions in terms of instrumentation and personnel.

## Conflict of interest

The authors Florian Sledz, Assegid M. Flatae and Mario Agio filed a patent for a diamond LED based on the results of this work (patent pending, DE102021123907.9).

## References

[1] S. Pezzagna and J. Meijer, Quantum computer based on color centers in diamond, Appl. Phys. Rev. **8**, 011308 (2021).

[2] J. L. O'Brien, Optical Quantum Computing, Science **318**, 1567 (2007).

[3] A. Beveratos, R. Brouri, T. Gacoin, A. Villing, J.-P. Poizat, and P. Grangier, Single Photon Quantum Cryptography, Phys. Rev. Lett. **89**, 187901 (2002).

[4] M. López, H. Hofer, and S. Kück, Detection efficiency calibration of single-photon silicon avalanche photodiodes traceable using double attenuator technique, J. Mod. Opt. **62**, 1732 (2015).

[5] B. Darquié, M. P. A. Jones, J. Dingjan, J. Beugnon, S. Bergamini, Y. Sortais, G. Messin, A. Browaeys, and P. Grangier, Controlled Single-Photon Emission from a Single Trapped Two-Level Atom, Science **309**, 454 (2005).

[6] D. B. Higginbottom, L. Slodička, G. Araneda, L. Lachman, R. Filip, M. Hennrich, and R. Blatt, Pure single photons from a trapped atom source, New J. Phys. **18**, 093038 (2016).

[7] H. G. Barros, A. Stute, T. E. Northup, C. Russo, P. O. Schmidt, and R. Blatt, Deterministic single-photon source from a single ion, New J. Phys. **11**, 103004 (2009).

[8] B. Lounis and W. E. Moerner, Single photons on demand from a single molecule at room temperature, Nature **407**, 491 (2000).

[9] X.-L. Chu, S. Götzinger, and V. Sandoghdar, A single molecule as a high-fidelity photon gun for producing intensity-squeezed light, Nat. Photonics **11**, 58 (2017).

[10] P. Lombardi, M. Trapuzzano, M. Colautti, G. Margheri, I. P. Degiovanni, M. López, S. Kück, and C. Toninelli, A Molecule-Based Single-Photon Source Applied in Quantum Radiometry, Adv. Quantum Technol. **3**, 1900083 (2020).

[11] I. Aharonovich, S. Castelletto, D. A. Simpson, C.-H. Su, A. D. Greentree, and S. Prawer, Diamond-based single-photon emitters, Rep. Prog. Phys. **74**, 076501 (2011).

[12] S. Castelletto and A. Boretti, Silicon carbide color centers for quantum applications, J. Phys. Photonics **2**, 022001 (2020).

[13] P. Michler, A. Kiraz, C. Becher, W. V. Schoenfeld, P. M. Petroff, L. Zhang, E. Hu, and A. Imamoglu, A Quantum Dot Single-Photon Turnstile Device, Science **290**, 2282 (2000).

[14] Z. Yuan, B. E. Kardynal, R. M. Stevenson, A. J. Shields, C. J. Lobo, K. Cooper, N. S. Beattie, D. A. Ritchie, and M. Pepper, Electrically Driven Single-Photon Source, Science **295**, 102 (2002).

[15] T. Heindel, C. Schneider, M. Lermer, S. H. Kwon, T. Braun, S. Reitzenstein, S. Höfling, M. Kamp, and A. Forchel, Electrically driven quantum dot-micropillar single photon source with 34% overall efficiency, Appl. Phys. Lett. **96**, 011107 (2010).

[16] D. J. P. Ellis, A. J. Bennett, S. J. Dewhurst, C. A. Nicoll, D. A. Ritchie, and A. J. Shields, Cavity-enhanced radiative emission rate in a single-photon-emitting diode operating at 0.5 GHz, New J. Phys. **10**, 043035 (2008).

[17] Z. Ge, T. Chung, Y.-M. He, M. Benyoucef, and Y. Huo, Polarized and Bright Telecom C-Band Single-Photon Source from InP-Based Quantum Dots Coupled to Elliptical Bragg Gratings, Nano Lett. **24**, 1746 (2024).

[18] S. Lagomarsino et al., Robust luminescence of the silicon-vacancy center in diamond at high temperatures, AIP Adv. **5**, 127117 (2015).

[19] A. Boretti, L. Rosa, A. Mackie, and S. Castelletto, Electrically Driven Quantum Light Sources, Adv. Opt. Mater. **3**, 1012 (2015).

[20] M. Petruzzella, F. M. Pagliano, Ž. Zobenica, S. Birindelli, M. Cotrufo, F. W. M. Van Otten, R. W. Van Der Heijden, and A. Fiore, Electrically driven quantum light emission in electromechanically tuneable photonic crystal cavities, Appl. Phys. Lett. **111**, 251101 (2017).

[21] Y. Yan and S. Wei, Doping asymmetry in wide-bandgap semiconductors: Origins and solutions, Phys. Status Solidi B **245**, 641 (2008).

[22] S. Koizumi and M. Suzuki, n-Type doping of diamond, Phys. Status Solidi A **203**, 3358 (2006).

[23] A. Lohrmann, S. Pezzagna, I. Dobrinets, P. Spinicelli, V. Jacques, J.-F. Roch, J. Meijer, and A. M. Zaitsev, Diamond based light-emitting diode for visible single-photon emission at room temperature, Appl. Phys. Lett. **99**, 251106 (2011).

[24] N. Mizuochi et al., Electrically driven single-photon source at room temperature in diamond, Nat. Photonics **6**, 299 (2012).

[25] B. Tegetmeyer, C. Schreyvogel, N. Lang, W. Müller-Sebert, D. Brink, and C. E. Nebel, Electroluminescence from silicon vacancy centers in diamond p–i–n diodes, Diam. Relat. Mater. **65**, 42 (2016).

[26] A. M. Berhane, S. Choi, H. Kato, T. Makino, N. Mizuochi, S. Yamasaki, and I. Aharonovich, Electrical excitation of silicon-vacancy centers in single crystal diamond, Appl. Phys. Lett. **106**, 171102 (2015).

[27] M. Haruyama et al., Electroluminescence of negatively charged single NV centers in diamond, Appl. Phys. Lett. **122**, 072101 (2023).

[28] D. Y. Fedyanin and M. Agio, Ultrabright single-photon source on diamond with electrical pumping at room and high temperatures, New J. Phys. **18**, 073012 (2016).

[29] I. A. Khramtsov, M. Agio, and D. Yu. Fedyanin, Dynamics of Single-Photon Emission from Electrically Pumped Color Centers, Phys. Rev. Appl. **8**, 024031 (2017).

[30] M. I. Current, Ion implantation of advanced silicon devices: Past, present and future, Mater. Sci. Semicond. Process. **62**, 13 (2017).

[31] R. Rouzbahani, K. J. Sankaran, P. Pobedinskas, and K. Haenen, Advances in *n*-Type Chemical Vapor Deposition Diamond Growth: Morphology and Dopant Control, Acc. Mater. Res. **5**, 775 (2024).

[32] J. Forneris et al., Electrical stimulation of non-classical photon emission from diamond color centers by means of sub-superficial graphitic electrodes, Sci. Rep. **5**, 15901 (2015).

[33] Y. Guo, W. Zhu, J. Zhao, S. Lin, Y. Yang, L. Lou, and G. Wang, Electroluminescence of NV by impact excitation and Stark shift in a MIM diamond structure, Appl. Phys. Lett. **119**, 252102 (2021).

[34] D. L. Scharfetter, Minority carrier injection and charge storage in epitaxial Schottky barrier diodes, Solid-State Electron. **8**, 299 (1965).

[35] S. Tiwari and D. J. Frank, Empirical fit to band discontinuities and barrier heights in III–V alloy systems, Appl. Phys. Lett. **60**, 630 (1992).

[36] E. H. Rhoderick and R. H. Williams, *Metal-Semiconductor Contacts*, 2nd ed (Clarendon Press ; Oxford University Press, Oxford [England] : New York, 1988).

[37] H. Kageshima and M. Kasu, Origin of Schottky Barrier Modification by Hydrogen on Diamond, Jpn. J. Appl. Phys. **48**, 111602 (2009).

[38] K. M. O'Donnell, M. T. Edmonds, A. Tadich, L. Thomsen, A. Stacey, A. Schenk, C. I. Pakes, and L. Ley, Extremely high negative electron affinity of diamond via magnesium adsorption, Phys. Rev. B **92**, 035303 (2015).

[39] W. M. H. Sachtler, G. J. H. Dorgelo, and A. A. Holscher, The work function of gold, Surf. Sci. **5**, 221 (1966).

[40] S. Koizumi, T. Teraji, and H. Kanda, Phosphorus-doped chemical vapor deposition of diamond, Diam. Relat. Mater. **9**, 935 (2000).

[41] A. M. Flatae et al., Silicon-vacancy color centers in phosphorus-doped diamond, Diam. Relat. Mater. **105**, 107797 (2020).

[42] R. Rouzbahani, S. S. Nicley, D. E. P. Vanpoucke, F. Lloret, P. Pobedinskas, D. Araujo, and K. Haenen, Impact of methane concentration on surface morphology and boron incorporation of heavily boron-doped single crystal diamond layers, Carbon **172**, 463 (2021).

[43] S. Lagomarsino et al., The center for production of single-photon emitters at the electrostatic-deflector line of the Tandem accelerator of LABEC (Florence), Nucl. Instrum. Methods Phys. Res. Sect. B Beam Interact. Mater. At. **422**, 31 (2018).

[44] F. Maier, J. Ristein, and L. Ley, Electron affinity of plasma-hydrogenated and chemically oxidized diamond (100) surfaces, Phys. Rev. B **64**, 165411 (2001).

[45] A. Di Bartolomeo, Graphene Schottky diodes: An experimental review of the rectifying graphene/semiconductor heterojunction, Phys. Rep. **606**, 1 (2016).

[46] M. Suzuki, S. Koizumi, M. Katagiri, T. Ono, N. Sakuma, H. Yoshida, T. Sakai, and S. Uchikoga, Electrical characteristics of n-type diamond Schottky diodes and metal/diamond interfaces, Phys. Status Solidi A **203**, 3128 (2006).

[47] H. Kato, M. Wolfer, C. Schreyvogel, M. Kunzer, W. Müller-Sebert, H. Obloh, S. Yamasaki, and C. Nebel, Tunable light emission from nitrogen-vacancy centers in single crystal diamond PIN diodes, Appl. Phys. Lett. **102**, 151101 (2013).

[48] A. M. Flatae et al., Single-photon emission from silicon-vacancy color centers in polycrystalline diamond membranes, Appl. Phys. Lett. **124**, 094001 (2024).

[49] H. Galal and M. Agio, Highly efficient light extraction and directional emission from large refractive-index materials with a planar Yagi-Uda antenna, Opt. Mater. Express **7**, 1634 (2017).

[50] H. Galal et al., Highly Efficient Light Extraction and Directional Emission from Diamond Color Centers Using Planar Yagi-Uda Antennas, arXiv:1905.03363.

# Electrical excitation of color centers in phosphorus-doped diamond Schottky diodes

Florian Sledz[1,*], Igor A. Khramtsov[2], Assegid M. Flatae[1], Stefano Lagomarsino[3], Silvio Sciortino[3,4,5], Shannon S. Nicley[6,7], Rozita Rouzbahani[6], Paulius Pobedinskas[6], Tianxiao Guo[8], Xin Jiang[8], Paul Kienitz[9], Peter Haring Bolivar[9], Ken Haenen[6], Dmitry Yu. Fedyanin[1,2] and Mario Agio[1,5]

[1]*Laboratory of Nano-Optics, University of Siegen, Germany*
[2]*Moscow Center for Advanced Studies, Moscow, Russia*
[3]*Istituto Nazionale di Fisica Nucleare, Sezione di Firenze, Italy*
[4]*Department of Physics and Astronomy, University of Florence, Italy*
[5]*National Institute of Optics (INO), National Research Council (CNR), Italy*
[6]*Institute for Materials Research (IMO) & IMOMEC, Hasselt University & IMEC vzw, Belgium*
[7]*Department of Electrical and Computer Engineering, Michigan State University, USA*
[8]*Lehrstuhl für Oberflächen- und Werkstofftechnologien, University of Siegen, Germany*
[9]*Group of Graphene-based Nanotechnology, University of Siegen, Germany*

*Corresponding author: Florian.Sledz@uni-siegen.de

**Hydrogen passivation on intrinsic diamond**

As mentioned in the main text, the recipe used for the hydrogen passivation was tested by the formation of a conductive layer on the surface of an intrinsic electronic grade diamond sample (EL SC Plate 4.5x4.5 mm, Element Six). Fig. S1a shows the IV-curve of the hydrogen passivated layer at the beginning of the measurement series shown in Fig. S1b and the curve measured after the measurement series and removal of the surface conductive layer by UV-ozone cleaning for 30 min. After removal the measured IV-curve overlaps with the parallel currents through the measurement devices. For the measurements the needle probes have been placed at the surface with a distance of ~20 µm.

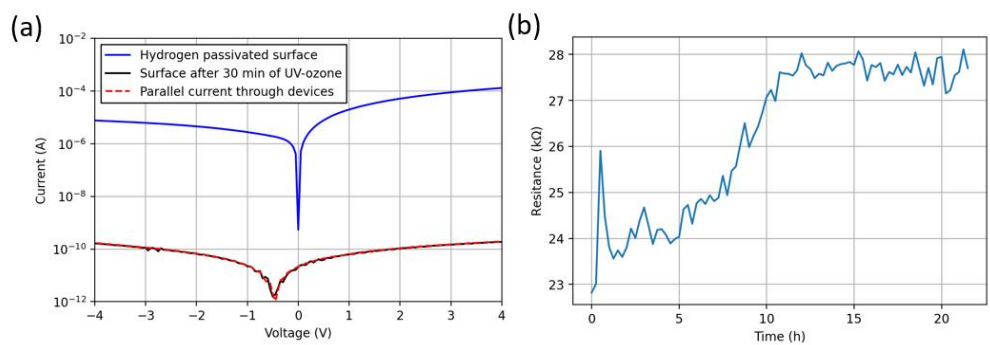

*Figure S1. (a) IV-curves of the hydrogen passivated intrinsic sample (blue), the same surface after 30 min of UV-ozone cleaning (black) and the parallel current through the measurement devices (red dashed). (b) Resistance of a hydrogen passivated layer measured every 15 min for 21 h.*

**Room temperature electroluminescence**

In the main text we presented the emission spectra at temperatures above 75 °C. For completeness, we are presenting here the electroluminescence spectra at 25 °C and at 50 °C in Fig. S2. The spectra at 25 °C are dominated by the emission of H3 color centers, while the emission of NV color centers can not be identified. As the temperature increases to 50 °C the zero-phonon line of NV color centers can be seen at ~578 nm on top of the falling slope of the H3 emission.

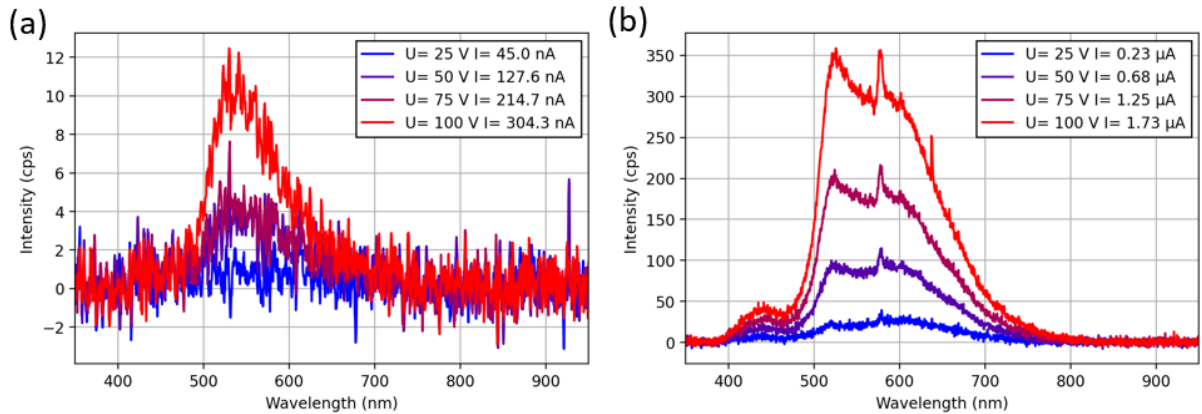

*Figure S2. Electroluminescence spectra at (a) 25 °C and (b) 50 °C. The spectra are dominated by the emission of H3 color centers. At 50 °C the emission of neutrally charged NV color centers can be seen on top of the falling slope of the H3 emission.*

**Current dependence of the emission**

Fig. 4b in the main text shows the dependence of the electroluminescence signal as a function of the voltage along with the measured current. According to Eq. 1 from the main text the emission rate depends on the charge carrier densities in the vicinity of the color center, which are proportional to the drift current densities at larger bias voltages. This results in a linear trend of the signal at larger voltages and the currents, respectively. Fig. S3a displays the electroluminescence signals in the main text in a linear scaling, while Fig. S3b shows the scaling of the signals at different distances as a function of current. However, as the carrier densities are not uniform, the slopes decrease with the distance.

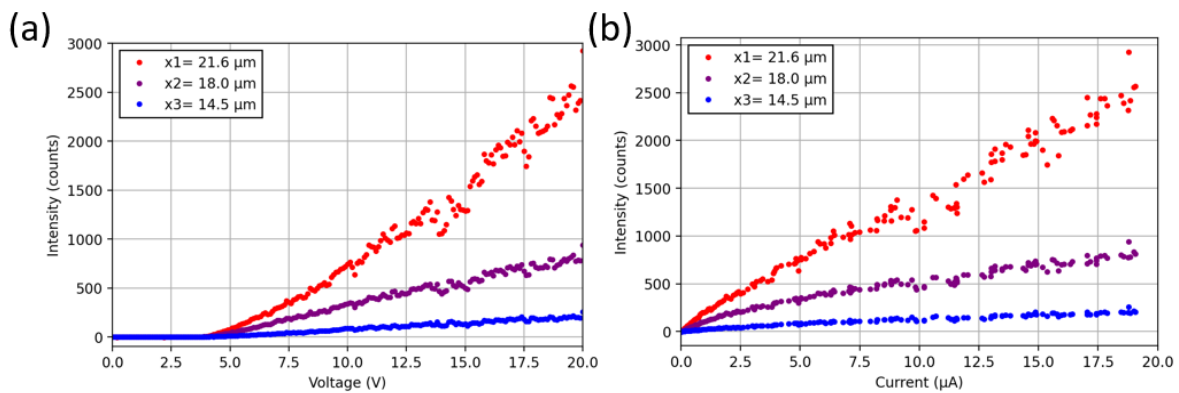

*Figure S3. (a) Electroluminescence signals at the three different positions shown in Fig. 4b in the main text in a linear scale. (b) The same signals as a function of current, measured at different positions. Towards larger voltages and currents the signals scale proportional to the respective values.*

**Electroluminescence signals in higher implanted regions**

In the main text the results of NV color centers in the region implanted with a silicon fluence of $10^{12}$ cm$^{-2}$ have been presented. Fig. S4 shows the comparison between the emission spectra obtained from the device presented in the main text with those at the center of the region implanted with a Si fluence of $10^{13}$ cm$^{-2}$. In the higher implanted regions, the ion-beam induced damage degrades the operation of the devices, which results in lower current densities and excitation rates. In Figs. S4c and S4d the emission of the H3 color centers dominates, while the emission of NV and SiV color centers can be barely seen at a temperature of 100 °C.

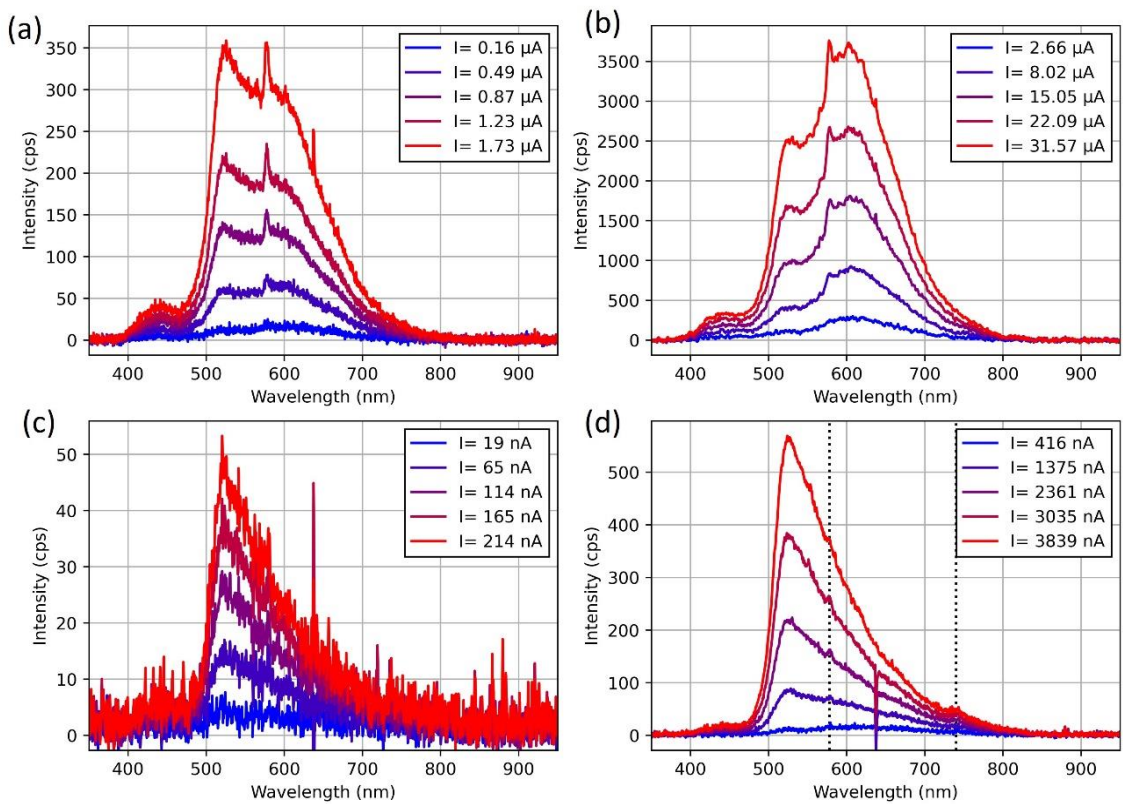

*Figure S4. Comparison of electroluminescence spectra in differently implanted regions (a,b) $10^{12}$ Si ions/cm² and (c,d) $10^{13}$ Si ions/cm² at increasing applied bias voltages of 20, 40, 60 80, and 100 V. The measurements are recorded at 50 °C (a,c) and 100 °C (b,d). (d) shows two small peaks at ~580 nm and ~740 nm corresponding to the zero-phonon lines of the NV$^0$ and SiV$^-$ color centers (black dotted), respectively.*

Fig. S5 shows the measurement results of a previous device generation in the region implanted with a Si ion-fluence of $10^{14}$ cm$^{-2}$ at a fixed bias voltage of 100 V and different temperatures. The spectra shown in Fig. S5a clearly show a zero-phonon line of SiV color centers around 740 nm on top of a broad nitrogen related background. The shift of the background to ~675 nm is due to the use of a different spectrometer (QEpro, OceanOptics) with a different spectral response compared to the spectrometer used in the main text. A Lorentzian peak with a linear background was fitted to the ZPL at different temperatures to determine its amplitude and the background beneath. The resulting signals are plotted in Fig. S5b as a function of temperature along the total measured current at 100 V. These trends clearly demonstrate the need for a higher sample purity as a spectral filtering or selective excitation of the SiV color centers is not possible under electrical excitation.

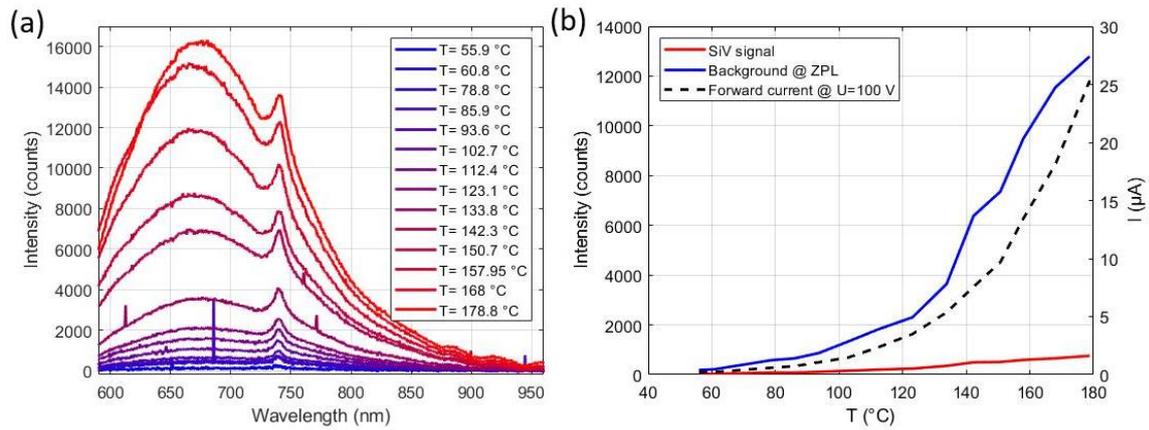

*Figure S5. (a) Electroluminescence spectra recorded at different temperatures and a forward bias voltage of U = 100 V in the region implanted with a Si fluence of $10^{14}$ cm$^{-2}$. In the spectra a clear ZPL of the SiV color centers is visible at ~740 nm on top of a wide background centered at ~675 nm. (b) Optical signals (left axis) and current (right axis) as a function of temperature at the same bias as in (a). The signal of the SiV color centers (red) is much smaller than the signal of the background (blue) beneath the ZPL.*

4